\documentclass[a4paper,12pt]{article}   
\usepackage{amsfonts}                   
\usepackage{amsmath}                    
\usepackage{mathtext}                   
\usepackage[dvips]{graphicx}            
\usepackage{color}
\usepackage{fleqn,epsf}
\usepackage{graphicx}

\def\shiftdown#1{#1\llap{\lower.04ex\hbox{#1}}}

\begin{document}

\begin{center}
{\large\bf On the  $\rho^o$-mesonM production in the inclusive
 proton-proton collision}
\end{center}
\vspace{1cm}
\noindent{\em A.I.\ Machavariani}

\vspace{3pt}
\noindent{ Joint\ Institute\ for\ Nuclear\ Research,\ LIT\ Dubna,\ Moscow\
region\ 141980,\ Russia\\
and High Energy Physics Institute of Tbilisi State University,
University str.  9 }\\
{\em Tbilisi 380086, Georgia }

\vspace{0.5cm}

\begin{abstract}

The production of the $\rho^o$- meson in the inclusive  proton-proton
scattering $p_A+p_B\to \rho^o+X$ is studied using
extension of parton model according to 
generalized vector meson dominance model (GVMD) 

in region of small $Q^2=m_{\rho^o}^2$ and
$\rho^o$-meson transverse momentum $\le 1-2GeV/c$. 
The realistic description of the experimental cross sections 
of $p_A+p_B\to \rho^o+X$ for
$4.9\le \sqrt{s}\le 65 GeV$ is achieved using an isotropic
distribution of the $\rho^o$-meson.

The resulting density matrix allows one to
select values of the quark masses, which lead to the isotropic
distribution of the emitted meson.
It is demonstrated that
the same cross sections of the reaction $p_A+p_B\to\rho^o +X$
for the isotropic distribution of the $\rho^o$-meson
can be obtained with  different sets of the quark masses 
and  corresponding coupling constants of 
the quark-meson vertex functions.


.
  
 \end{abstract}

\newpage
\centerline{\bf  Introduction}

\vspace{0.25cm}

Nowadays  the transverse momentum dependent parton distribution
function (TMD PDF) model is successfully applied
for the  description of the lepton pair production
in numerous reactions with momentum transfer $\sqrt{Q^2}\geq 2.4 GeV$
\cite{TMDCollins2015,Angeles-Martinez,Nachtmann2015,Lu1,Lu2}.
Moreover, in \cite{Goloskokov1,Goloskokov2} detailed description
of the production of the  $\rho.\phi$ mesons in the exclusive 
electron-proton  scattering is achieved
in the framework of the generalized parton distributions (GPD)
with the simple Gaussian functions for the distributions of the
transverse quark momentum's in the quark-meson wave functions.
  The fitting of the unintegrated pion multiplicities in
the semi inclusive
deep inelastic scattering (SIDIS) reactions $\ell N\to\ell'hX$
was fulfilled in the work of Torino group \cite{Anselmino3}
  using corresponding TMD PDF model.
These successful applications of the TMD PDF model 
rely on existence of the factorization theorem 
\cite{TMDCollins2015,CollinsBook2011,Rogers,Bacchetta,Anselmino1},
under which the cross sections of the considered reactions are 
a product of the cross sections of  parton annihilation  
process and the nucleon TMD  PDFs.
This condition was proved
when $Q^2>> {\bf k_T}^2$ \cite{TMDCollins2015,CollinsBook2011},
where ${\bf k_T}$ denotes the  transverse momentum of the
final observed particle.
For production of the $\rho$ meson
with transverse momentum ${\bf k_T}^2\leq 1(GeV/c):2$
in the  the inclusive reactions 
this condition is not fulfilled  because
$\sqrt{Q^2}=m_{\rho}=0.775 GeV/c$ and $Q^2\sim {\bf k_T}^2$.
The calculations  \cite{Goloskokov1,Goloskokov2} and \cite{Anselmino3}
  of reactions   $\ell p\to\ell'\rho(\phi)p'$ and $\ell p\to\ell'hX$
were performed within different TMD PDF models and different 
realization of the factorization theorem. 

Other kind of successful description of the  
production of the light and vector mesons, $\Delta^{++,+,o}$ and $\Lambda$-resonances
and proton and antiproton in the inclusive proton-proton
collision at $\sqrt{s}=27.5GeV$\cite{Anguilar91} 
with $p_T^2\le 4(GeV/c)^2$ 
was obtained  within the FRITIOF model 
\cite{FRITIOF1,FRITIOF2} which
 based upon semiclassical  approach and string dynamics.
This description was obtained using parton model and corresponding 
standard PDF functions without restrictions coming from the
spin of the particles and factorization theorem.
Various applications of  FRITIOF model  
for the  pp, pA and AA collisions  within parton model 
are given in refs. \cite{HIJING1,HIJING2,HIJING3}.

On the other hand existence of the transition  $\gamma^*\longleftrightarrow V$
allows one simply connect the cross sections of the reactions 
$p_A+p_B\to \gamma^*X$ and  $p_A+p_B\to V X$
in the framework of the parton model 
(see e.g. ch. 14.3.3 \cite{Leader}) based on  the GVMD 
(generalized vector meson dominance) model 
\cite{Bugaev,Schildknecht}-\cite{Likhoded,Sullivan76}. 
Moreover GVMD model  \cite{Schildknecht,Kwienski,Sullivan76}
imply  an extension of the PDF into region of
very small $x$ and $Q^2<1GeV^2$ (or $Q^2=m_V^2$) and large
 $p_T^2$. Within GVMD model was achieved 
 explanation and good description  of the various experimental data 
of deep inelastic scattering (DIS) in the region $Q^2\le 1GeV^2$ and small $x$.
An other way of extension of PDF
in the region of the very small $x$ and 
$Q^2<1GeV^2$  is given in \cite{Glueck1998,Glueck2008} for
ultra cosmic ray and neutrino astronomy.

\vspace{0.05cm}
\begin{figure}[htb]
\includegraphics[width=10.5 cm]{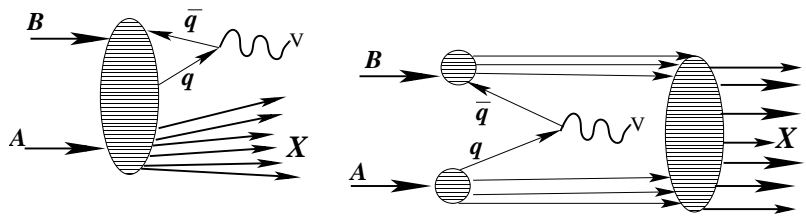}
\vspace{-0.5 cm}
\caption { {\it
({\bf a}) Full amplitude of the 
reaction $p_A+p_B\to\rho^+X$ and its representation
in the parton fusion model ({\bf b}).
}}
\end{figure}


The purpose of this work is  to examine validity of 
 extension of the usual fusion parton model
\cite{Halsen,Leader} in the framework of the GVMD model
for description  of the $\rho^o$-meson production  
in the inclusive proton-proton scattering
$p_A+p_B\to \rho^o+X\to 1+2+X$ ($1,2\equiv \pi^+\pi^-\ or\ ee^+$).  
In this approach the full amplitude of the reaction
$p_A+p_B\to V+X$ in Fig. 1a is replaced by the parton fusion 
diagram   in Fig. 1b
and in addition transverse momentum of partons  is taken into account 
via the Gaussian multiplier in PDF.
We will use the  fusion model 
\cite{Ter2}, where quarks and mesons
are on the energy and mass shell.

The article consists of four Sections. 
The first Section deals with  the relationship 
between cross 
sections of the reactions $p_A+p_B\to \gamma^*+X$ and 
$p_A+p_B\to V+X$ according to GVMD model. 
Unlike to the usual fusion model we take into account 
 the vector and tensor coupling parts in the  quark-meson vertex. 
An analytic expression for the density matrix of the reaction 
$p_A+p_B\to \rho^o+X$
and extraction of the isotropic distributions of the $\rho^o$-meson
are given in the Sect. 2. 
The numerical results
 for the  cross sections of the reaction $p_A+p_B\to \rho^o+X$
 are presented in Sect. 3. 
These calculations were performed for the isotropic distributions 
when the nondiagonal elements of the density matrix are equal to zero 
and diagonal elements are equal to $1/3$.
A brief conclusion is given in Sect. 4. 

\vspace{5mm}

\begin{center}
  {\bf 1. 
Vector meson dominance model (VMD) and  cross section of the reaction  $p_A+p_B\to V+X$}
\end{center}

\vspace{5mm}

We shall describe $\rho$-meson production in the 
 inclusive proton-proton scattering
 $$p_A({\bf P_A}S_A)+p_B({\bf P_B}S_B)\to \rho({\bf k}M)+X 
\eqno(1.1)$$
where  ${\bf k}$ and $M$ denote a momentum and magnetic 
quantum number of  the $\rho$-meson, ${\bf P_A}S_A$ and 
${\bf P_B}S_B$ stand for the momentum and magnetic quantum 
number of the  proton $A$ and $B$ correspondingly.
In particular, in the c.m. frame of the protons $A$ and $B$  
 $P=({\bf P_A})_Z=-({\bf P_B})_Z$ and  
${\bf k}$ is placed in the $ZX$-plane.


In this paper
the differential cross section of the reaction 
$p_A+p_B-\rho^oX-\ell\ell^+X$ (or  $p_A+p_B-\rho^oX-\pi\pi X$)
is determined according to extension of the original Drell-Yan 
parton model for the dilepton photo production reaction 
$p_A+p_B-\gamma^*X-\ell\ell^+X$. This extension is performed 
within VMD. In particular, VMD model implies
replacement of the quark-antiquark  annihilation
vertices $q{\overline q}-\gamma^*$ with the   
vertices $q{\overline q}-\rho^o$. This procedure 
enables to replace the non-resonant Drell-Yan  cross section 

$${ { d{\sigma}^{MM'}_{p_A+p_B\to \gamma^*X}}\over{d{ {\bf k_T}^2}dy} }=
{{m_N^2}\over{ (4\pi)^2Ps^{1/2}}} {1\over {3(2b^2)}}
\sum_{n=u,d} \biggl[
{ {2 {\sf m}_1{\sf m}_2}\over {m_{V}^2}}
\int d^2{\bf q_1}_T\int d^2{\bf q_2}_T 
\delta({\bf k}_T-{\bf q_1}_T-{\bf q_2}_T)$$
$$\Bigl({\sf f}_{n/A}(x_1,{\bf q_1}_T)
{\sf f}_{{\overline n}/B }(x_2,{\bf q_2}_T)
+(1\longleftrightarrow 2)\Bigr)\biggr]
\Gamma^{MM'}_{n{\overline n}-\gamma^*}
\eqno(1.2a)$$

with the resonant cross section


$${ { d{\sigma}^{MM'}_{p_A+p_B\to VX}}\over{d{ {\bf k_T}^2}dy} }=
{{m_N^2}\over{ (4\pi)^2Ps^{1/2}}} {1\over {6(2b^2)}}
\sum_{n=u,d} \biggl[
{ {2 {\sf m}_1{\sf m}_2}\over {m_{V}^2}}
\int d^2{\bf q_1}_T\int d^2{\bf q_2}_T 
\delta({\bf k}_T-{\bf q_1}_T-{\bf q_2}_T)$$
$$\Bigl({\sf f}_{n/A}(x_1,{\bf q_1}_T)
{\sf f}_{{\overline n}/B }(x_2,{\bf q_2}_T)+(1\longleftrightarrow 2)\Bigr)\biggr]
\Gamma^{MM'}_{n{\overline n}-V},
\eqno(1.2b)$$
where ${\bf k_T}$ and $y=1/2ln\Big[(k_o+k_Z)/(k_o-k_Z)\Bigr]$ 
denote the transverse 
momentum and  the longitudinal
rapidity of the $\gamma^*$ or $\rho$-meson.
${\bf q}$, $s$ and $n$ stand for a momentum, 
$Z$-projection of the spin and flavour $n=u,d$ of the quarks.
The factor $m_N^2/((4\pi)^2Ps^{1/2})$ arose in 
(1.2a,b) due to replacement of the  variables ${\bf k}$
with ${\bf k_T}^2$ and $y$
in the standard differential cross sections \cite{IZ}, 
 the multiplier $1/3$ arose in (1.2a,b) from 
averaging over the color quantum numbers of the quarks
\cite{Halsen,Leader}, an additional $1/2$ in (1.2b)
is used according to the isotopic structure
of the $\rho^o$-meson  $|\rho^o>=(|u{\overline u}>-|d{\overline d}>)/\sqrt{2}$.
$\Gamma^{MM'}_{n{\overline n}-\gamma^*}$ and
$\Gamma^{MM'}_{n{\overline n}-V}$ are the product of the $\gamma^*-q{\overline q}$
 or  $V-q{\overline q}$ vertices in (1.2a,b). In the both cases
$\Gamma^{MM'}_{n{\overline n}-{\cal V}}$ (${\cal V}\equiv \gamma^*,V$)
have the same form
$$\Gamma^{MM'}_{n{\overline n}-{\cal V}}
=\xi^{\mu}_{\cal V}({\bf k},M){\xi^{\nu}}^*_{\cal V}({\bf k},M')
<{\bf q}_1s_1,n;{\bf q}_2s_2,{\overline n}|J_{\cal V\mu}(0)|0>
<0|J_{\cal V\nu}(0)|{\bf q}_1s_1,n;{\bf q}_2s_2,{\overline n}>,
\eqno(1.3a)$$
where 
$\xi^{\mu}_{\cal V}({\bf k},M)$ is the polarization function of 
 $\gamma^*$ or $V$-meson and $J_{\gamma^*\mu}(x)=\Box\Phi_{\gamma^*}(x)$
and $J_{ V\mu}(x)=(\Box+m_V^2)\Phi_V(x)$ are the corresponding source operators.

The cross sections (1.2a,b) are a part of the 
cross section of the Drell-Yan reaction $p_a+p_B-\ell\ell^+X$. 
Therefore $\Gamma^{MM'}_{n{\overline n}-{\cal V}}$ (1.3a)  is 
ingredient of the cross section 
of  the  reaction $q{\overline q}-\gamma^*-\ell\ell^+$ or  $q{\overline q}-V-\ell\ell^+ $ 
$$\sigma^{{\tilde M}{\tilde M}',MM'}_{n{\overline n}-{\cal V}-\ell\ell^+}={{4\pi}\over 3}
 \Gamma^{{\tilde M}{\tilde M}'}_{n{\overline n}-{\cal V}}{1\over{(p_{\ell}+p_{\ell^+})^2-m_{\cal V}^2}}\Gamma^{ MM'}_{{\cal V}-\ell\ell^+},
\eqno(1.3b)$$
where $m_{\gamma^*}^2=0$ and $m_{V}$ is the mass of the $V$-meson. 
$p_{\ell}$ and $p_{\ell^+}$ are four-momentum's of the final
electrons and $\Gamma^{ MM'}_{{\cal V}-\ell\ell^+}$ is defined 
through  product of vertices $\ell\ell^+ -{\cal V}$ in analogy with (1.3a).

Unlike to the original Drell-Yan reaction \cite{DYorig},
 (1.2a) contains integration over the 
transverse momentum's (see e.g. \cite{Collins2014}).
Validity of  replacement of the intermediate photons 
with the vector meson  follows from existence of the  
transition  $\gamma^*\longleftrightarrow V$
and it was tested many times in  VMD models
\cite{Bugaev,Schildknecht}-\cite{Sullivan76}.


We take PDF ${\sf f}_{n/A}(x,{\bf q}_T)$ 
in (1.2a,b) same form as in TMD  PDF model   

$${\sf f}_{n/A}(x,{\bf q}_T)=
f_{n/A}(x){ {e^{-{\bf q}_T^2/(2 b^2)}}\over{2 b^2} };\ \ \ \ \ \ 
{\sf f}_{{\overline n}/B}(x,{\bf q}_T)=
f_{{\overline n}/B)}(x){ {e^{-{\bf q}_T^2/(2 b^2)}}\over{2 b^2}},
\eqno(1.4)$$
where ${\bf q_T}$ denote the
transverse momentum of the quark. 

$$x_1={{k_o+k_{Z}}\over {2P}};\ \ \ \ \ x_2={{k_o-k_{Z}}\over {2P}},
\eqno(1.5)$$
where $k^o=\sqrt{m_V^2+{\bf k_T}^2+k_Z^2}$.

The same Gaussian functions
of transverse momentum distribution were used in 
\cite{Goloskokov1,Anselmino3}.

The factor $(2 {\sf m}_1{\sf m}_2)/ m_{V}^2$ in (1.2a,b)
contain  ${\sf m}_1{\sf m}_2$
which arose in  the intermediate integration over the quarks
with ${\sf m}_1d^3{\bf q_1}/q_1^o$
and ${\sf m}_2d^3{\bf q_2}/q_2^o$ according to\cite{IZ}.
This factor cancel with the corresponding expression in (2.6c).
The factor $1/(2b^2)$ in (1.2a,b) insures the correct dimension
of these cross sections.

In (1.3a) the vertex $q{\overline q}-V$ is

$$<{\bf q}_1s_1,n;{\bf q}_2s_2,{\overline n}|J_{\mu}(0)|0>=
g_V^{n{\overline n}}{\overline v}({\bf q_2})\gamma_{\mu}u({\bf q_1})
+g_T^{n{\overline n}}{\overline v}({\bf q_2})
{{i\sigma_{\mu\nu} (q_1+q_2)^{\nu}}\over{{\sf m}_1+{\sf m}_2}}
u({\bf q_1}),\eqno(1.6)$$
where $u({\bf q_2})$, ${\overline v}({\bf q_2})$,
$\gamma_{\mu}$, $\sigma_{\mu\nu}$ are well-known spinors and
 Dirac matrices \cite{IZ}.

$<{\bf q}_1s_1,n;{\bf q}_2s_2,{\overline n}|J_{\mu}(0)|0>$ (1.6)
is considered on  mass shell of the $V$-meson, i.e. 
$(q_1+q_2)^2=m_V^2$.
Therefore, the form factors  $g_V^{n{\overline n}}$ and 
$g_T^{n{\overline n}}$ are determined by corresponding coupling constants.


$g_V^{n{\overline n}}$ and $g_T^{n{\overline n}}$ can be 
determined through
the similar electromagnetic constants of the 
$n{\overline n}-\gamma^*$-system in the same way as
the coupling constants of the  
$\rho NN$-vertex are constructed via the coupling constants  
of the $\gamma^* NN$-vertex within the GVMD model 
\cite{Sullivan76,Schildknecht,Bugaev}, where
$$ (g_V^{\rho NN})^2=G_V^2e^2;\ \ \ \ \ \ \ \ \ \ \ \ \ \ \
(g_T^{\rho NN})^2=G_T^2e^2.\eqno(1.7)$$
In analogy  to (1.7),
 $g_V^{n{\overline n}}$ and $g_T^{n{\overline n}}$ are determined
through the charges of quarks $e_n$ and antiquarks 
$e_{\overline n}$: 

$$(g_V^{n{\overline n}})^2=g_V^2e_n^2;\ \ \ \ \ \ \ \ \ \ 
(g_T^{n{\overline n}})^2=g_T^2e_n^2,\eqno(1.8)$$
or 

$$g_V^{u{\overline u}}=g_Ve_u;\ \ \ \ \ g_T^{u{\overline u}}=g_Te_u;\ \ \ \ \ 
g_V^{d{\overline d}}=g_Ve_d;\ \ \ \ \ g_T^{d{\overline d}}=g_Te_d,\eqno(1.9)$$
where $e_u=2/3$ and 
$e_d=-1/3$.



\vspace{5mm}
\begin{center}
{\bf 2. The density matrix and isotropic distribution}
\end{center}

\vspace{5mm}
In order to construct the density matrix from the cross sections
(1.2b), we consider this cross section in the rest frame
of the $\rho$-meson (Fig. 3), where 
${\bf k^{\star}}=0$, ${\bf q_1^{\star}=-q_2^{\star}}$ 
$${\bf q}^{\star}_i=
{\bf q_i}+ {{ {\bf k}}\over{ m_{V}}}
\biggl[{ { ( {\bf k q_i})}\over {m_{V}+k^o}}-q_i^o\biggr];
\ \ \ \ \ \ \ \ \ \ \ \ \ \ \ \ \ \ i=1,2.
\eqno(2.1)$$

In this paper meson and quarks in the $V$-meson-quark 
vertex (1.6) are on mass and energy shell. Then from the
energy-momentum conservation in the meson rest frame
we have  $m_V={q_{1}^*}^o+{q_{2}^*}^o$ and 
${\bf q_1}^{\star}+{\bf q_2}^{\star}=0$. These relations enables
determine $|{\bf q}^{\star}|^2$ and ${q_{1,2}^*}^o$ 
through the masses of quarks and meson
$$|{\bf q}^{\star}|^2={ {\Bigl(m_{V}^2-({\sf m}_1+{\sf m}_2)^2\Bigr)
\Bigl(m_{V}^2-({\sf m}_1-{\sf m}_2)^2\Bigr)}
\over {4m_{V}^2} };\ \ \ 
{q_{1,2}^*}^o={{m_V^2\pm{\sf m}_1^2
\mp{\sf m}_2^2}\over {2m_V}}.
\eqno(2.2)$$ 

Therefore, 
it is convenient to use the  spherical coordinates 

$$ 
{\bf q}^{\star}\equiv {\bf q}^{\star}_1= |{\bf q}^{\star}|
\Bigl(sin\alpha^{\star} cos\beta^{\star},
sin\alpha^{\star}\sin\beta^{\star},
\cos\alpha^{\star}\Bigr),\eqno(2.3)$$

where
$$cos\alpha^*= {{{\bf q}_{1Z}}\over {|{\bf q}^{\star}|}}-
{{{\bf k}_{Z}}\over {|{\bf q}^{\star}|}}\ 
{{q_1^o+{q_1^*}^o}\over {m_V+k^o}};\ \ \ \ \ 
tg\beta^*={ { {\bf q_1}_Y}\over 
{ {\bf q_1}_X-{\bf k}_{X}{{{q_1^o+{q_1^*}^o}\over {m_V+k^o}} }} }.
\eqno(2.4)$$

\vspace{0.5cm}
\begin{figure}[htb]
\includegraphics[width=6.5 cm]{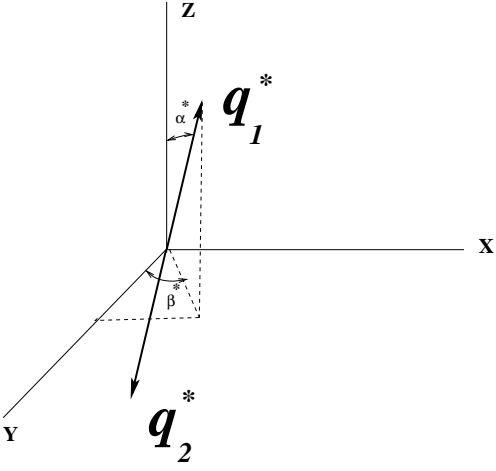}
\caption { {\it
${\bf q_1^*}$ and ${\bf q_2^*}$ (2.1) and the angles $\alpha^{\star}$
and $\beta^{\star}$ (2.3) in the rest frame of the $V$-meson.}}
\end{figure}
\vspace{0.5cm}

A direct generalization of the Gordon identities  for the  
particles with  unequal masses ${\sf m}_1$ and ${\sf m}_2$

$${\overline v}({\bf q^{\star}_2})
{{i\sigma_{\mu\nu} (q^{\star}_1+q^{\star}_2)^{\nu}}\over{{\sf m}_1+{\sf m}_2}}
u({\bf q^{\star}_1})=-{\overline v}({\bf q^{\star}_2})\gamma_{\mu}u({\bf q^{\star}_1})
+{\overline v}({\bf q^{\star}_2})
{ {(q_1^{\star}-q_2^{\star})_{\mu}}\over{{\sf m}_1+{\sf m}_2}}u({\bf q^{\star}_1})
\eqno(2.5)$$
allows one to represent  (1.3) as 
$$\sigma^{MM'}_{n,{\overline n} }=e_n^2{\Sigma}^{MM'}\eqno(2.6a)$$
$${ \Sigma}^{MM'}=\xi^{\mu}(0,M){\xi^{\nu}}^{*}(0,M')Tr
\biggl[(g_V-g_T)^2\gamma_{\mu}{ { (q_1^*\gamma)+{\sf m}_1}\over{2{\sf m}_1}}
\gamma_{\nu}{ { (q_2^*\gamma)-{\sf m}_2}\over{2{\sf m}_2}}$$
$$+g_T^2{ { (q_1^*-q_2^*)_{\mu}}\over{{\sf m}_1+{\sf m}_2}}
{ { (q_1^*-q_2^*)_{\nu}}\over{{\sf m}_1+{\sf m}_2} }
{ { (q_1^*\gamma)+{\sf m}_1}\over{2{\sf m}_1}}
{ { (q_2^*\gamma)-{\sf m}_2}\over{2{\sf m}_2}}$$
$$+g_T(g_V-g_T)\Bigl(\gamma_{\mu}{ { (q_1^*-q_2^*)_{\nu}}\over{{\sf m}_1+{\sf m}_2} }
{ { (q_1^*\gamma)+{\sf m}_1}\over{2{\sf m}_1}}
{ { (q_2^*\gamma)-{\sf m}_2}\over{2{\sf m}_2}}
+{ { (q_1^*-q_2^*)_{\mu}}\over{{\sf m}_1+{\sf m}_2}}
{ { (q_1^*\gamma)+{\sf m}_1}\over{2{\sf m}_1}}\gamma_{\nu}
{ { (q_2^*\gamma)-{\sf m}_2}\over{2{\sf m}_2}}
\Bigr)\biggr],\eqno(2.6b)$$
where $(q^*\gamma)\equiv q_{\sigma}^*\gamma^{\sigma}$.

Simple calculation
 leads to the following expression for 
${ \Sigma}^{MM'}$
$${ {2 {\sf m}_1{\sf m}_2}\over {m_{V}^2}}
{\Sigma}^{MM'}=(g_V-g_T)^2{{m_V^2-({\sf m}_1-{\sf m}_2)^2}
\over{m_V^2}}\biggl[{\delta}^{MM'}+\delta Re\Bigl({\cal A}^{MM'}\Bigr)\biggr],
\eqno(2.6c)$$
where factor ${ {2 {\sf m}_1{\sf m}_2}\over {m_{V}^2}}$
is taken from (1.2a,b),
${\delta}^{MM'}=1$ if $M=M'$ and 
${\delta}^{MM'}=0$ if $M\ne M'$,
$$
{\cal A}^{MM'}=
\begin{bmatrix}
{{\sin^{2}(\alpha^*)}\over{2}} & -{{\sin(2\alpha^*)}\over{2\sqrt{2}}}e^{-i\beta^*} 
& -{{\sin^{2}(\alpha^*)}\over{2}}e^{2i\beta^*} \\
-{{\sin(2\alpha^*)}\over{2\sqrt{2}}}e^{i\beta^*} & \cos^{2}(\alpha^*) 
& {{\sin(2\alpha^*)}\over{2\sqrt{2}}}e^{i\beta^*} \\
-{{\sin^{2}(\alpha^*)}\over{2}}e^{-2i\beta^*} & {{\sin(2\alpha^*)}\over{2\sqrt{2}}}e^{-i\beta^*} & {{\sin^{2}(\alpha^*)}\over {2}}
\end{bmatrix},
\eqno(2.6 d)$$

$$\delta={ { m_{V}^2-({\sf m}_1+{\sf m}_2)^2}\over
{m_{V}^2} }
\biggl[{{g_T^2}\over {(g_V-g_T)^2}} 
{ { m_{V}^2-({\sf m}_1+{\sf m}_2)^2}\over
{( {\sf m}_1+{\sf m}_2)^2}} -{{2g_T}\over {(g_V-g_T)}}-1
\biggr],\eqno(2.7)$$

Afterwards the cross section (1.2b) takes the form 
$$
{{d\sigma^{MM'}_{p_A+p_B\to VX}}\over{d{\bf k}_X^2dy}}=
{1\over {6(4\pi)^2}}\ {{m_N^2}\over{Ps^{1/2}}}
{{1}\over{2b^2}}
{\sf d}<{\Sigma}^{MM'}>,\eqno(2.8)$$
where
$$<{\Sigma}^{MM'}>={1\over {\sf d}}\int d^2{\bf q_{2T}}
\Bigl(e_u^2{\sf f}_{u/A}(x_1,{\bf k_X-q_{2T}})
{\sf f}_{{\overline u}/B}(x_2,{\bf q_{2T}})$$
$$+e_d^2{\sf f}_{d/A}(x_1,{\bf k_X-q_{2T}}){\sf f}_{{\overline d}/B}(x_2,{\bf q_{2T}})
+(1\longleftrightarrow 2)\Bigr){ \Sigma}^{MM'}
\eqno(2.9a)$$

$${\sf d}= \int d^2{\bf q_{2T}}
\Bigl(e_u^2{\sf f}_{u/A}(x_1,{\bf k_X-q_{2T}})
{\sf f}_{{\overline u}/B}(x_2,{\bf q_{2T}})$$
$$+e_d^2{\sf f}_{d/A}(x_1,{\bf k_X-q_{2T}}){\sf f}_{{\overline d}/B}(x_2,{\bf q_{2T}})
+(1\longleftrightarrow 2)\Bigr)\eqno(2.9 b)$$

and  $<{\cal F}(\alpha^{\star},\beta^{\star})>$ denotes the average value of
a function ${\cal F}(\alpha^{\star},\beta^{\star})$ 

$$<{\cal F}(\alpha^{\star},\beta^{\star})>={1\over {\sf d}} \int d^2{\bf q_{2T}}
\Bigl(e_u^2{\sf f}_{u/A}(x_1,{\bf k_X-q_{2T}}){\sf f}_{{\overline u}/B}(x_2,{\bf q_{2T}})$$
$$+e_d^2{\sf f}_{d/A}(x_1,{\bf k_X-q_{2T}}){\sf f}_{{\overline d}/B}(x_2,{\bf q_{2T}})
+(1\longleftrightarrow 2)\Bigr) {\cal F}(\alpha^{\star},\beta^{\star}).\eqno(2.9 c)$$

The elements of the density matrix $\rho_{MM'}$ are defined as
$$\rho_{MM'}={ { {{ d\sigma^{MM'}_{p_A+p_B\to VX}}/{d{\bf k}_X^2dy}} }
\over{{{d\sigma_{p_A+p_B\to VX}}/{d{\bf k}_X^2dy}} }};
\ \ \ \ \ \ \ \ \ \ \ \ \ \ \ \ \ \ \ \ \ \ 
 {{d\sigma_{p_A+p_B\to VX}}\over{d{\bf k}_X^2dy}}=
\sum_M {{d\sigma^{MM}_{p_A+p_B\to VX}}\over{d{\bf k}_X^2dy}}.
\eqno(2.10)$$

Using formulas (2.8) and (2.9 a,b,c,d), we can obtain $\rho_{MM'}$ 
(2.10) explicitly 

$$\rho_{00}={ {\delta<cos^2(\alpha^{\star})>+1}\over {\delta+3} };\ \ \ \ \ 
\rho_{11}=\rho_{-1-1}={1\over 2}{ {\delta<sin^2(\alpha^{\star})>+2}\over {\delta+3} }
\eqno(2.11 a)$$
$$Re\rho_{1-1}={1\over 2}<sin^2(\alpha^{\star})cos(2\beta^{\star})>
{ {\delta}\over {\delta+3} };\ \ \ \ \ 
Re\rho_{10}={1\over {2\sqrt{2}}}<sin(2\alpha^{\star})cos(\beta^{\star})>{ {\delta}\over {\delta+3} }
\eqno(2.11 b)$$

$$Im\rho_{1-1}={1\over 2}<sin^2(\alpha^{\star})sin(2\beta^{\star})>
{ {\delta}\over {\delta+3} };\ \ \ \ \ 
Im\rho_{10}={1\over {2\sqrt{2}}}<sin(2\alpha^{\star})sin(\beta^{\star})>{ {\delta}\over {\delta+3} }.
\eqno(2.11c)$$

From the density matrix $\rho_{MM'}$ (2.11a,b,c),
one can  extract 
 conditions for  the isotropic distribution  when
$$\rho_{oo}=\rho_{11}=\rho_{-1-1}=1/3;\ \ \ \rho_{M\ne M'}=0. \eqno(2.12)$$
In particular, if
$$\delta=0, \eqno(2.13)$$
then the conditions (2.12) are fulfilled.
 
Afterwards, using a special choice  of the masses 
of  quarks ${\sf m}_1$ and ${\sf m}_2$`
in  $\delta$ (2.7), one can satisfy
the condition (2.13). There exist two  possibilities: 

\begin{enumerate}
\item  ${\sf m}_1$ and ${\sf m}_2$ are at threshold 
$|{\bf q^*}|^2=0$, where 
$${\sf m}_1+{\sf m}_2=m_{V}\ \ \ \ \ or\ \ \ {\sf m}_1={\sf m}_2=m_{V}/2
.\eqno(2.14 a)$$
 
\item $g_T/g_v$ satisfy the  equation 
$${{g_T^2}\over {(g_V-g_T)^2}} 
{ { m_{V}^2-({\sf m}_1+{\sf m}_2)^2}\over
{( {\sf m}_1+{\sf m}_2)^2 }}
-{{2g_T}\over {(g_V-g_T)}}-1=0.
\eqno(2.14b)$$

\end{enumerate}

The solution of the equation (2.14b) determine
$g_T/g_V$ and $(g_V-g_F)^2$ 
through  ${\sf m}_1$, ${\sf m}_2$ and $ m_V$. 
  
$$g_T= g_V
{{ {\sf m}_1+{\sf m}_2}\over{ m_V}}\ \ \ 
which\ \ gives\ \ \ 
(g_V- g_T)^2=g_V^2
{ { (m_V -{\sf m}_1-{\sf m}_2)^2}
\over{m_V^2}}.\eqno(2.14c)$$
The solution (2.14c)  for the small (current) quark masses
${\sf m}_{1,2}\simeq 5-10MeV$ gives the small values of $g_T$.
However, $g_T/({\sf m}_1+{\sf m}_2)=g_V/m_V$ in the vertex  (1.6)
is not  negligible.
Thus the isotropic distribution with the current quark masses 
produces redefinition  
$g_T/({\sf m}_1+{\sf m}_2)\longrightarrow g_V/m_V$ 
in the quark-meson vertex (1.6).

Using the condition (2.13), one can rewrite 
the cross section (2.8) as  

$${{d\sigma_{p_A+p_B\to VX}}\over{d{\bf k}_X^2dy}}=
{{(g_V-g_T)^2}\over {2(4\pi)^2}}{{m_N^2}\over{Ps^{1/2}}}
{{1}\over{2b^2}}\int {\bf d q_{2T}^2}
\Bigl(e_u^2{\sf f}_{u/A}(x_1,{\bf k_X- q_{2T}})
     {\sf f}_{{\overline u}/B}(x_2,{\bf q_{2T}})$$
     $$+e_d^2{\sf f}_{d/A}(x_1,{\bf k_X- q_{2T}})
{\sf f}_{{\overline d}/B}(x_2,{\bf q_{2T}}) + (1\longleftrightarrow 2)\Bigr).\eqno(2.15)$$
This expression depends on ${\sf m}_1$ and ${\sf m}_2$ through
$(g_V-g_T)^2$ only. It is easy to see that the different sets
of $g_V$, $g_T$, ${\sf m}_1$ and ${\sf m}_2$ can give the same 
$(g_V-g_T)^2$ i.e. the same cross section (2.15).

In particular, for the current quark masses
${\sf m}_{1,2}\simeq 5-10MeV$
the cross section (2.15) remains the same
after replacement  $(g_V-g_T)^2\Longrightarrow g_V^2(current)$.

An other way to get  the equal diagonal elements of the isotropic
density matrix
$$\rho^{00}=\rho^{-1-1}=\rho^{11}=1/3 \eqno(2.16 a)$$
turns out if 
$$<{cos^2}(\alpha^{\star})>={1\over 2} <{sin^2}(\alpha^{\star})>={1\over 3}.\eqno(2.17)$$

In this case the nondiagonal elements of the density matrix are
 not equal to zero
$$\rho^{M\ne M'}\ne 0 \eqno(2.16 b)$$
and the distribution is anisotropic.
  
\vspace{10mm} 

\centerline{\bf 3. Numerical calculations of cross sections}
\vspace{10mm}

We have numerically estimated the  cross sections 
${{d\sigma_{p_A+p_B\to \rho^oX}}/{d{\bf k}_X^2}}$,
 ${{d\sigma_{p_A+p_B\to \rho^oX}}/{dy}}$ and 
$\sigma_{p_A+p_B\to \rho^oX}$
based upon   
${{d\sigma_{p_A+p_B\to \rho^oX}}/{d{\bf k}_X^2dy}}$ (2.15) 
for the isotropic distribution (2.12).

The  parameter  $B\equiv 1/(2b^2)$ of the transverse part of the 
 PDF (1.4) was chosen as $B_1=3.6(GeV/c)^{-2}$ ($b=0.4082GeV/c$) and 
 $B_2=3.0(GeV/c)^{-2}$ ($b=0.3727GeV/c$).
The corresponding curves in Figs. 3,4 and 5 are displayed with 
 solid and dotted lines.
The value of $B=3.6\pm 0.4(GeV/c)^{-2}$
was used in \cite{Bloebel74} to describe  experimental data at 
a momentum of the incoming proton 
$12GeV/c$ and $24GeV/c$.  
$B=3.3\pm 0.2(GeV/c)^{-2}$ was need in \cite{Albrow79} 
to reproduce the  experimental cross sections in the region
 $23.6\le\sqrt{s}\le 63.0 GeV$ and 
$B=2.59\pm 0.1(GeV/c)^{-2}$ was used in \cite{Anguilar91}.
The magnitude of the parameter $B=2.6-3.6(GeV/c)^2$ in (1.4) 
differs largely from the same parameter  $B=0.5-0.7(GeV/c)^2$
in the TMD PDF model for fitting of the SIDIS unintegrated pion
multiplicities \cite{Anselmino3} in the reaction $\ell N\to\ell'hX$ in the
kinematical region $Q^2>1.69(GeV)^2$ and $0.2 GeV/c<{\bf k}_T<0.9GeV/c$.


\begin{figure}[htb]
\includegraphics[angle=-90,width=14.5 cm]{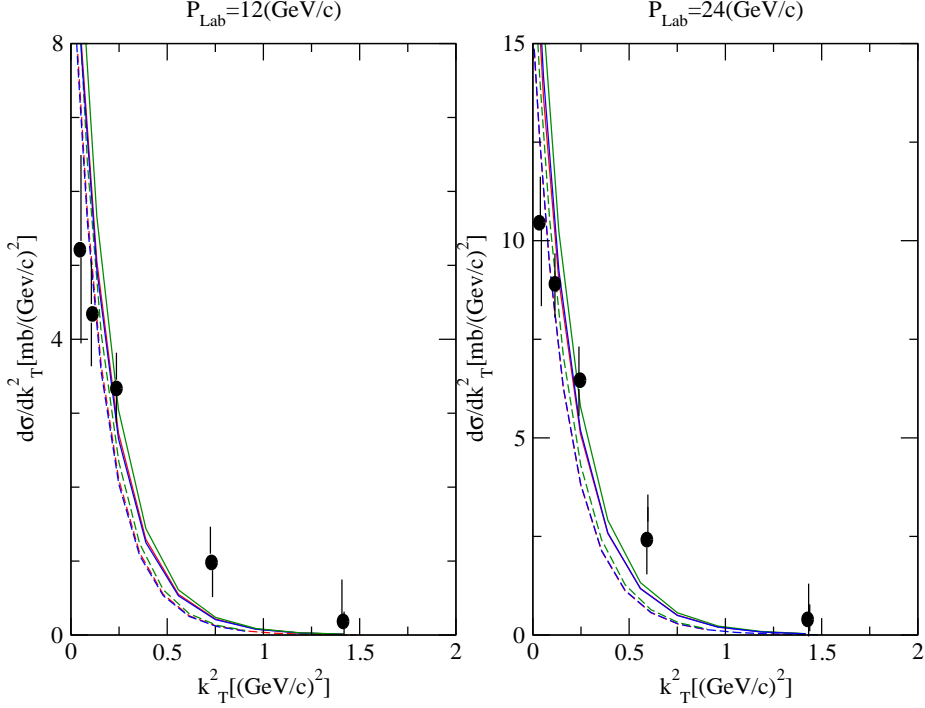}
\vspace{-0.5 cm}
\caption { {\it 
The distribution over the
 transverse momentum of the $\rho^o$-meson
$k^2_X$  at $P_{Lab}=12GeV/c$ and 
$24 GeV/c$. Experimental data from \cite{Bloebel74}.
[M08], [G98] and [G08] indicate curves with the PDF from 
\cite{Martin,Glueck1998} and \cite{Glueck2008} correspondingly. 
}}
\end{figure}

Adjusted parameter $(g_T-g_V)^2/(4\pi)$ is  strongly correlated with
$B$. For $B=3.6(GeV/c)^{-2}$ we take $(g_T-g_V)^2/(4\pi)=32.1$
and for $B=3.0(GeV/c)^{-2}$ we use $(g_V-g_T)^2/(4\pi)=32.1*1.6$.
These values are consistent with $g_V^2/(4\pi)$ and $g_T^2/(4\pi)$ 
within the VMD model\cite{O'Connell} and the VMD 
model for the constituent quarks \cite{Weber},
where $g^{\rho NN}_T/g^{\rho NN}_V=5\ or\ 6.6\pm 0.6$ and 
$2\le (g^{\rho NN}_V)^2/(4\pi)\le 3$ $(g^{\rho NN}_V\equiv g_V)$. By virtue of (2.14c), 
for the current quark masses $\simeq 5-10MeV$
one obtains the cross section (2.15) if 
$g_V^2(current)\simeq (g_V-g_T)^2$.

We have used three models of the standard PDF $f_{n/A}(x)$ 
\cite{Martin,Glueck2008,Glueck1998} which are depicted 
in Figs. 3,4 and 5 with red, green and blue curves, respectively.
In \cite{Martin} PDF are defined in regions $Q^2>1(GeV)^2$
$x>0.001$, or $x>0.01$,
or $x>0.1$  for the different reactions and different quarks.
In \cite{Glueck1998} PDF are constructed in the regions $Q^2>1(GeV)^2$
and $x>0.001$ and this region is extended for in $x>10^{-5}$ for the 
ultra cosmic ray and neutrino astronomy. The extension of  
$f_{n/A}(x)$ in $Q^2<1 (GeV)^2$ and in the ultra small $x>10^{-9}$ region is given  \cite{Glueck2008}.

${{d\sigma_{p_A+p_B\to \rho^oX}}/{d{\bf k}_X^2}}$
for the different parameters $B$ and corresponding $(g_V-g_T)^2$,
and different PDF models  are depicted in Fig. 3. 
These curves are rather close to each other and
describe well the 
experimental data \cite{Bloebel74} 
excluding the region $0.5\le {\bf k}_X^2\le 0.75(GeV/c)^2$.
Unlike  \cite{Bloebel74}, in
${{d\sigma_{p_A+p_B\to\rho^o X}}/{d{\bf k}_X^2}}$ (1.2) and (2.15) one can 
not extract exactly the dependence on ${\bf k}_X^2$
via factor $e^{-B{\bf k}_X^2}$ because $x_1$ and $x_2$ (1.5) depend on
 ${\bf k}_X^2$ through $k_o=\sqrt{m_V^2+{\bf k}_X^2+{\bf k}_Z^2}$.
Similarly, unlike \cite{Halsen,Leader},  
$d\sigma_{p_A+p_B\to VX}/dy$ contains integration over ${\bf k}_X^2$
of  $e_u^2{f}_{u/A}(x_1){ f}_{{\overline u}/B}(x_2)
    +e_d^2{f}_{d/A}(x_1){ f}_{{\overline d}/B}(x_2)$.

\vspace{5.5 mm}

\begin{figure}[htb]
\includegraphics[width=14.5 cm]{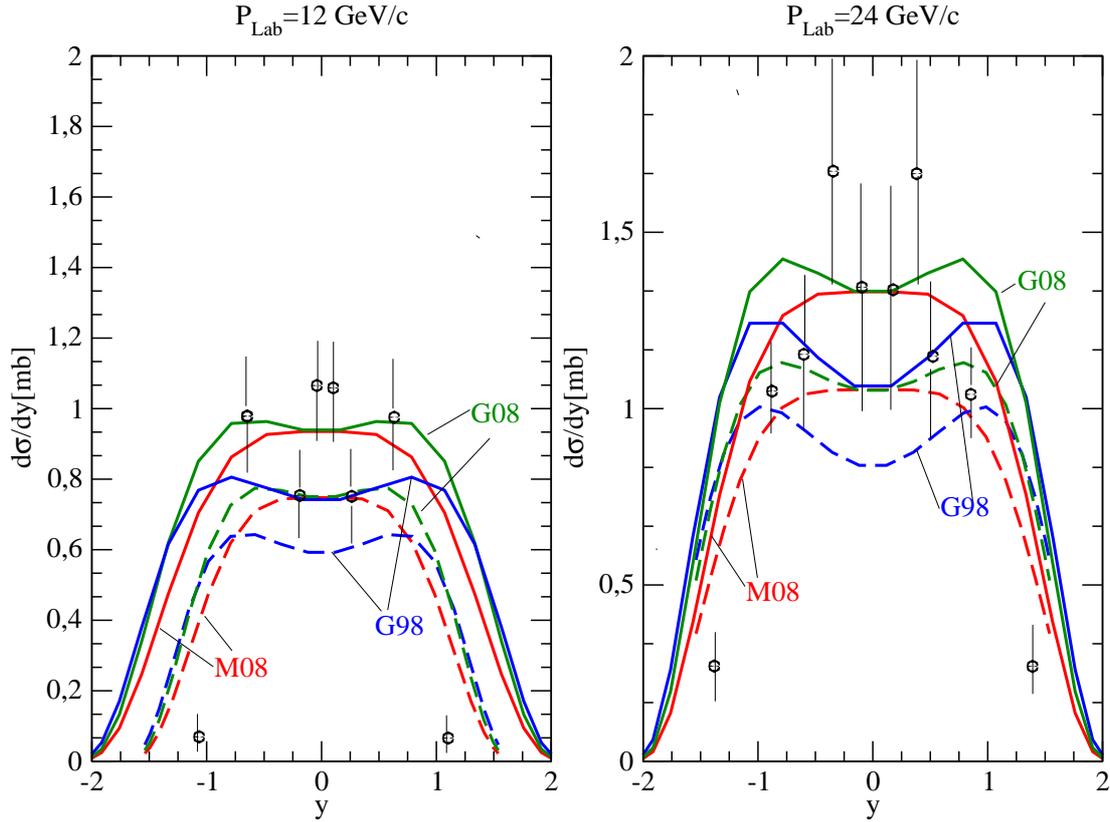}
\caption { {\it
The same as in Fig. 3, but for the 
distribution over the longitudinal rapidity $y$.}}
\end{figure}
\vspace{10.5 mm}

 $d\sigma_{p_A+p_B\to \rho^oX}/dy$ in Fig. 4  have a  
typical behavior for small $y$ close to the constant, followed by a 
 rapid drop. Similar behavior of these cross sections
was also obtained in \cite{Sullivan76}
using an additional scaling condition 
between the cross sections of the 
light vector mesons and pions. 
In contrast to the distribution of 
${\bf k}_X^2$ in Fig. 3, the distribution of $y$ in Fig. 4 
significantly depend on the choice of the
PDF and $2b^2$ parameter  in (1.4).

The qualitative description of the experimental 
cross sections 
${{d\sigma_{p_A+p_B\to hX}}/{d{\bf k}_X^2}}$ and  
$d\sigma_{p_A+p_B\to hX}/dy$ for  production of the mesons 
$\pi^{\pm,o},\rho^{\pm,o},\omega,\eta,f^o,K^{\pm},
K^{*o},{\overline K}^{o*},\phi$, $\Delta^{++,+,o}$ and $\Lambda$-resonances and proton 
and antiproton
in the inclusive proton-proton
collision at $\sqrt{s}=27.5GeV$\cite{Anguilar91}
  was obtained  within the FRITIOF model 
\cite{FRITIOF1,FRITIOF2} within semiclassical  approach and string dynamics. 
Relationship between 
Monte Carlo models for multiple jet production 
in pp, pA and AA collisions  and parton model 
is given in refs. \cite{HIJING1,HIJING2,HIJING3}

\newpage

\begin{figure}[htb]
\includegraphics[width=18.5 cm]{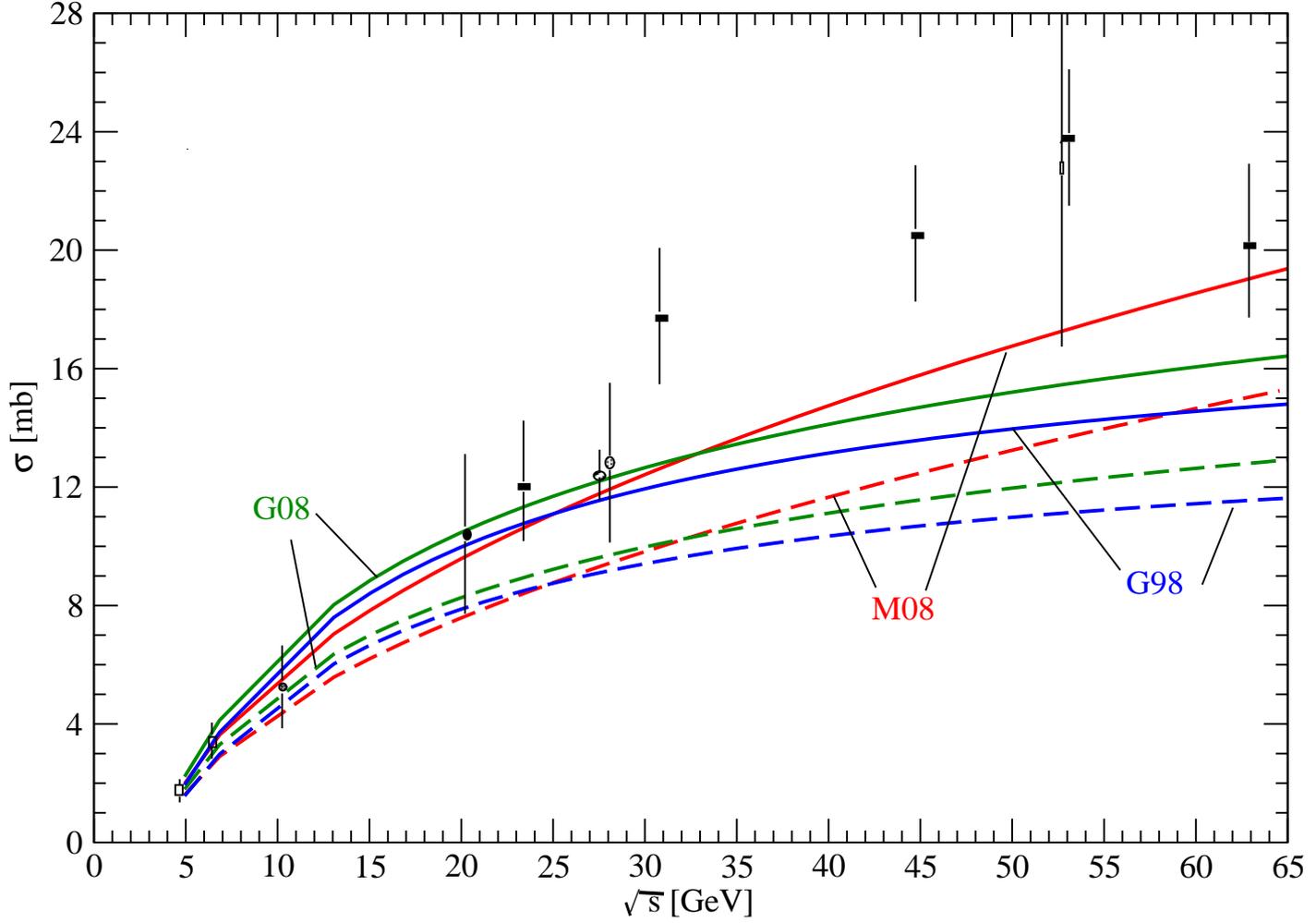}
\caption { {\it Same as in the Fig. 3 and 4, but for the
total  cross section as a function of 
$\sqrt{s}$. Experimental data at 4.93, 6.84 GeV from \cite{Bloebel74},
at 1.2 GeV from \cite{Ammosov76},at 26.8 GeV from \cite{Kiohmi78,Albrow79},
at 23.6,\ 30.6,\ 44.6,\ 52.8,\ 63.0 GeV from \cite{Albrow79}at 52.5 GeV from \cite{Drijard81}, 
and at 27.5 GeV from \cite{Anguilar91}.}}
\end{figure}
\vspace{0.25 mm}

Total cross sections $\sigma_{p_A+p_B\to \rho^oX}$ in Fig. 5
 also significantly depend on the choice of
PDF and parameters $2b^2$ in (1.4) and  $(g_T-g_V)^2$.
However, these cross-sections have important features: 
1) They increase dramatically 
in the energy region up to $\sqrt{s}\sim 20GeV$. In particular,
$$ {{\sigma_{p_A+p_B-\rho^o X}(P_{beam}=24Gev/c)}\over
{\sigma_{p_A+p_B-\rho^o X}(P_{beam}=12Gev/c)}}\simeq 2\eqno(3.1)$$
2) In the energy region $\sqrt{s}> 20GeV$ these cross sections 
are growing much slower than before 
$\sqrt{s}< 20GeV$.
In the considered  model, this behavior of 
$\sigma_{p_A+p_B\to \rho^oX}$ is 
determined by the standard PDF.
 The dependence of $\sigma_{p_A+p_B\to \rho^oX}$
on choice of PDF
parameters $2b^2$  in (1.4) and related  $(g_V-g_N)^2$
is much stronger in the energy region $\sqrt{s}> 20GeV$.



\newpage
{\bf 4.\ \ \ Conclusion }

\vspace{0.25 cm}

We have shown that the parton model
with an isotropic distribution of the $\rho^o$ meson (2.12)
reproduces  realistically the
experimental  cross sections of the inclusive 
proton-proton scattering $p_A+p_B\to \rho^0 X$.
This is consistent with the experimental results \cite{Bloebel74}, 
  according which
the angular distributions of this reaction  at $P_{lab}=12\ and\ 24GeV/c$
 are roughly isotropic.

Present formulation of the reaction  $p_A+p_B\to \rho^o+X$ 
 based on extension of the  original parton model
for the Drell-Yan reaction $p_A+p_B\to \gamma*+X$ 
within GVMD\cite{Bugaev,Schildknecht}-\cite{Likhoded,Sullivan76}
which requires continuation of PDF in the region of the small 
$x$ and $Q^2<1GeV^2$. For this aim  we have used three different PDF 
\cite{Glueck1998,Glueck2008,Martin}, where   
 for $|x|<<1$ and $Q^2<1GeV^2$ PDF are constructed exactly
\cite{Glueck1998,Glueck2008} using the corresponding experimental 
data. Cross sections in Fig. 4,5 depend significantly 
on the choice of the PDF, but different PDF give qualitatively
 similar results.  The isotropic cross section (2.15) strongly depends 
on the choice  of the parameters $2b^2$ of 
 PDF (1.4) and the corresponding parameter $(g_V-g_T)^2$. 
Sensitivity on the correlated parameters $2b^2$
and $(g_V-g_T)^2$ of the cross sections increases 
in the region $20GeV<\sqrt{s}<65GeV$.

The  cross section (2.15) for the isotropic 
distributions contains only two adjustable 
parameters: $2b^2$ in  the standard PDF (1.4) 
and the related constants $(g_V-g_T)^2$ 
from the $\rho NN$ vertex (1.6).
The considered cross section (2.15) depends on the quark masses 
through $(g_V-g_T)^2$ as it follows from  
(2.14a,b) and (2.14c) for the isotropic distribution (2.12).
This allows one to get the several set
of the different quark masses
which yield  the same cross section (2.15)
with the same $(g_V-g_T)^2$. In particular,
we have demonstrated that one can obtain the same isotropic
distributions within
 the constituent and current quark models.
The detailed theoretical and experimental investigation
of the anisotropic contributions in the  considered 
cross sections and the density matrix
$\rho_{MM'}$ (2.10)-(2.11a,b,c) 
will allow one to estimate the mechanisms of  removing  
this degeneracy.





Author thanks O.V.Teryaev for useful remarks and discussions.


\vspace{0.1cm}

\begin{thebibliography} \\
\bibitem{TMDCollins2015} J. Collins and T. Roggers,Phys. Rev. {\bf D91}(2015)074020;
arXiv:1412.3820v3[hep-ph](2015)
\bibitem{Angeles-Martinez} R.Angeles-Martinez et al,arXiv:1507.95267v1(2015) 
y
\bibitem{Nachtmann2015} O. Nachtmann. Ann. of Phys., {\bf 350}(2014)347
\bibitem{Lu1} Zhun Lu and I. Schmidt, Phys Rev.  {\bf D81}(2010) 03423.
\bibitem{Lu2} Bing Zhang, Zhun Lu, Bo-Qhuang Ma and I. Schmidt, Phys Rev.  {\bf D77}
(2008) 054011.
\bibitem{Goloskokov1} S.V. Goloskokov and P. Kroll. Eur. J.Phys. {\bf C42}(2005)146.
\bibitem{Goloskokov2} S.V. Goloskokov and P. Kroll. Eur. J.Phys. {\bf C50}(2014)281.
\bibitem{Anselmino3} M. Anselmino et al, JHEP.  {\bf 04}(2014) 005.

\bibitem{CollinsBook2011} J. Collins. Foundations of perturbative QCD (CUP, 2011)
\bibitem{Rogers} T.C.Rogers and P.J.Mulders, Phys. Rev. {\bf D81}(2010) 094006.
\bibitem{Bacchetta} A. Bacchetta et al, JHEP.  {\bf 02}(2007) 093.
\bibitem{Anselmino1} M. Anselmino et al, Phys. Rev. {\bf D83}(2011) 114019.
\bibitem{Halsen} F. Halsen and A.D. Martin. 
Quarks and Leptons. An Introduction Course in Modern Particle 
Physics.
x(John Wiley and Sons, New York-Chechester-Brisbane-Toronto-Singapure)
 1984.
\bibitem{Leader} E. Leader and E Predazzi. 
An Introduction to 
Gauge Theories and "New Physics".
(Cembridge University Press, London, New York, New Rochelle, 
Melbourne and  
and Sydney) 1982.

\bibitem{Anguilar91} M. Anguilar-Bonitez  at al,.Z. Phys. {\bf C50}(1991)405.

\bibitem{FRITIOF1} B. Andersson, G. Gustefsou and 
B. Nilsson-Almqvist., Nucl. Phys.  {\bf B281}(1987) 
 289.
\bibitem{FRITIOF2} V. Uzhinsky.  Archiv: 1404, 2026[hep-ph] (2014).
\bibitem{HIJING1} X.N. Wang and M. Gyulassy. Phys. Rev. {\bf D44} (1991) 3501.
\bibitem{HIJING2} X.N. Wang. Phys. Rep. {\bf 280} (1997) 287.
\bibitem{HIJING3} W.T, Dong, X.N. Wang and R. Xu. Phys. Rev. {\bf C83} (2011) 014915.

\bibitem{Collins2014} J. Collins, 
Int. J. Mod. Phys. Conf. Series{\bf 25}(2014) 1160001;

\bibitem{Bugaev} E.V. Bugaev and B.V. Mangazeev, Phys. Rev.{\bf D90}(2014)014026-1
\bibitem{Schildknecht} D. Schildknecht,Acta  Phys. Pol. {\bf B37}(2006)595
\bibitem{Kwienski} J. Kwienski and B. Badejek, Z. Phys. {\bf C43}(1989)253
\bibitem{Likhoded} A.K. Likhoded, S.B. Slabostitskii
 and A. M. Tolstikov. Sov. J. Nucl. Phys, {\bf 35} (1982) 1240.
\bibitem{Sullivan76} S. Chavin and J. D. Sullivan, Phys. Rev. D 19, D 11 (1976)



\bibitem{Bloebel74} V. Blobel at al,. Phys. Lett. {\bf B58}(1974)83.
\bibitem{Ammosov76} V.V. Ammosov at al,. Sov. J. Nucl. Phys. {\bf 24}(1978)30. 
\bibitem{Kiohmi78} H. Kiohmi at al,. Contr.in 19th Int.Conf. hIgh-en phys. Tokyo 1978
\bibitem{Albrow79} M.G.Albrow at al,. Nucl. Phys. {\bf B155}(1979)39.
\bibitem{Drijard81} D. Drijard at al,. Z. Phys. {\bf C9}(1981)293.

\bibitem{IZ} C.  Itzykson  and C. Zuber.
Quantum Field theory. (New York, McGraw-Hill) 1980.
\bibitem{Ter2} A.V. Efremov and O. V. Teryaev. 
JINR Preprint P2-82-832, 1982 (in russian).

\bibitem{DYorig} S. D. Drell and T.-M. Yan.
Phys. Rev. Lett. {\bf 25} (1970) 316.
\bibitem{O'Connell} H.B. O'Connell, A.G. Williams, M. Bracco and G.Krein, Phys.Lett.  {\bf B370}(1995)12.
\bibitem{Weber} H.J. Weber, Phys.Lett.  {\bf B233}(1989)267.
\bibitem{Glueck1998} M. Glueck, E. Reya and A. Vogt., 
Eur. Phys J.  {\bf C5}(1998) 461.
\bibitem{Glueck2008} M. Glueck, P. Jimenez-Toledo and                   E. Reya., Eur. Phys J.  {\bf C53}(2008) 355.
\bibitem{Martin} A. D. Martin, W. J. Stirling, R.S.Thorne,,
and G. Watt. Eur. Phys J.  {\bf C63}(2009) 189.
\end{thebibliography}
\end{document}